\documentclass[twocolumn,showpacs,preprintnumbers,amsmath,amssymb]{revtex4}
\usepackage{amssymb}
\usepackage{color}
\usepackage{graphicx}
\usepackage{dcolumn}
\usepackage{bm}

\begin{document}

\title{Laser driving of Superradiant scattering at variable incidence angle}

\author{Bo Lu}
\author{Xiaoji Zhou}\thanks{Electronic address: xjzhou@pku.edu.cn}
\author{Thibault Vogt}
\author{Zhen Fang}
\author{Xuzong Chen}

\affiliation{School of Electronics Engineering and Computer Science, Peking University, Beijing 100871, People's Republic of China}

\date{\today}

\begin{abstract}
We study superradiant scattering from a Bose-Einstein condensate using a pump laser incident at variable angle and show the presence of asymmetrically populated scattering modes. Experimental data reveal that the direction of the pump laser plays a significant role in the formation of this asymmetry, result which is in good agreement with numerical simulations based on coupled Maxwell-Schr\"{o}dinger equations. Our study complements the gap of previous work in which the pump laser was only applied along the short axis or the long axis of a condensate, and extends our knowledge about cooperative scattering processes.
\end{abstract}

\pacs{03.75.Kk, 42.50.Gy, 42.50.Ct, 32.80.Qk}

\maketitle

\section{INTRODUCTION}
Since the trailblazing experiment of superradiant Rayleigh scattering from a Bose-Einstein condensate (BEC) \cite{ketterle1999}, numerous experiments and theoretical developments have emerged in cooperative scattering of light and atoms in ultracold atomic gases concerning Rayleigh superradiant scattering \cite{ketterle1999,ketterle2003,Moore,You,Piovella,puhan,Zobay,twof1,yang,twof3}, amplification of matter waves \cite{inouye1999mwa,kozuma1999mwa,inouye2}, coherent atomic recoil lasing (CARL) \cite{slama,zhou}, Raman superradiance \cite{Dominik,Yutaka,Mary} and its application in probing the spatial coherence of an ultracold gas \cite{raman1,raman2}. Why matter-wave superradiance can occur only when the pump laser is red detuned has been explained recently \cite{detuning1,detuning2}. These interaction processes of light and matter are important to the fields of cold atom physics, cold molecular physics, quantum optics, and quantum information science.

In a typical BEC superradiance experiment, a cigar-shaped BEC is illuminated by a laser beam along its short axis or long axis. Because of the large optical density along the long axis, the spontaneous emission of BEC atoms excited by the pump laser is amplified along this axis. This creates the superradiant radiation modes that are the so-called end-fire modes. In lowest order, forward peaks are generated when a BEC atom absorbs a photon from the pump laser and emits a photon into an end-fire mode. In contrast, backward peaks are produced when a BEC atom absorbs a photon from an end-fire mode and emits a photon into the pump laser. The atoms which undergo the process of absorption and emission form superradiant scattering patterns and the recoiling atoms have well-defined momenta. In the short-pulse regime, the relevant time scale is so short compared to the recoil period that the energy mismatch is overcome by a large energy uncertainty. Therefore, both forward and backward modes are populated. In the long-pulse regime, in contrast, the pump pulse is long enough that backward scattering is suppressed and only forward modes are populated. If the pump laser is incident along the short axis of a BEC, X-shape or fan-like patterns are observed \cite{ketterle2003} and the two end-fire modes are symmetrically propagating in opposite directions along the long axis. If the pump laser is incident along the long axis of a BEC, all recoiling atoms distribute in line and this is called the line-pattern \cite{ljt}. In this geometry, only one end-fire mode propagating backward with respect to the pump laser dominates the superradiance process and the two end-fire modes are asymmetric. Several experimental observations and theoretical investigations describing the above two cases have been published. Up to now, no asymmetric scattering pattern has been reported for other angular configuration. The main purpose of the present paper is to resolve this interesting question.

In this paper, we study asymmetric scattering from a Bose-Einstein condensate. First we present the experimental observations of angular superradiant scattering from a BEC with the laser pulse incident at different angles. The change of the scattering pattern with the angle, between symmetric and asymmetric superradiance, is demonstrated. Then taking into account the angular configuration, we use the semiclassical Maxwell-Schr\"{o}dinger equations to describe the coupled dynamics of matter-wave and optical fields. The calculated atomic recoiling patterns and numerical simulations of end-fire modes are consistent with our experimental results. This study complements previous reported results, especially the gap of understanding between the two kinds of experiments in which the pump laser was only applied along the short axis or the long axis of a BEC, and extends our knowledge about cooperative scattering processes.

\section{EXPERIMENTAL DESCRIPTION AND RESULTS}
The experimental setup is similar to the one reported in our previous work \cite{yang,twof3}. After laser cooling and evaporation cooling, a cigar-shaped BEC with $2\times 10^{5}$ $^{87}$Rb atoms is created in the $|F=2, m_{F}=2\rangle$ hyperfine ground state in the quadrupole and Ioffe configuration trap with a length of $L\simeq100\ \mathrm{\mu m}$ and a diameter of $W\simeq10\ \mathrm{\mu m}$. To induce superradiant scattering, the cigar-shaped BEC is illuminated with a single off-resonant pump laser beam that is red-detuned by 1.45 GHz from the $5 S_{1/2}, F=2 \rightarrow 5 P_{3/2}, F^{'} =3$ transition. The linearly polarized pump laser is incident at a variable angle $\theta$ with respect to the long axis of the BEC in the $\mathbf{\hat{x}}$-$\mathbf{\hat{z}}$ plane, and its electric field vector is along the $\mathbf{\hat{y}}$ axis, see Fig. 1(a). We carefully avoid the presence of the light back-reflected by the vacuum cell windows and other optical devices. In order to probe the momentum distribution of atoms, after switching off the magnetic trap and applying the laser pulse, we image the ballistically expanding cloud after 30 ms time of flight using resonant probe light propagating along the $\mathbf{\hat{y}}$ axis.

\begin{figure}[htbp]
\centering
\includegraphics[width=8.5cm]{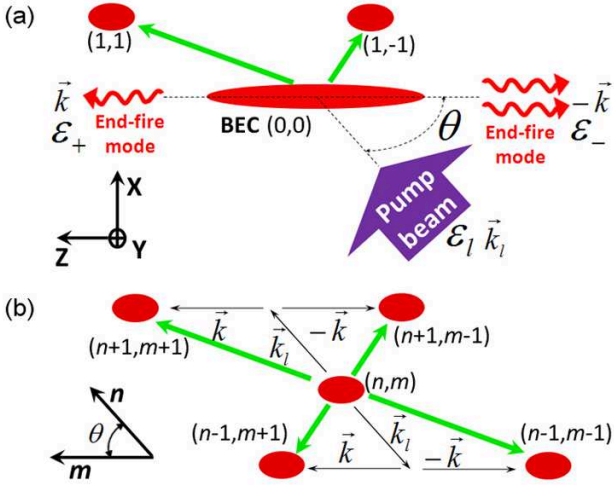}
\caption{(Color online) Schematic drawing of the angular superradiant scattering. (a) A cigar-shaped BEC is illuminated by a single pump beam (wave vector\ $\vec{k}_{l}$) linearly polarized along the $\mathbf{\hat{y}}$ axis and propagating at an angle of $\theta$ with respect to the long axis of the BEC. The emission of two end-fire modes $\mathcal{E}_{+}(\vec{k})$ and $\mathcal{E}_{-}(-\vec{k})$ are supposed to occur along the axis of the condensate. Only two first-order modes are shown. (b) The atomic side modes are denoted in momentum space. Each mode is labeled with a pair of integers $(n,m)$ which represents the momentum $\hbar(n\vec{k}_{l}+m\vec{k})$ possessed by the scattering atoms in this side mode. With this notation, the side mode (0, 0) describes the condensate at rest. A forward-scattering event transfers an atom from mode $(n,m)$ to mode $(n+1,m\pm1)$ and a backward event transfers one to mode $(n-1,m\pm1)$.}
\end{figure}

\smallskip

We have checked first that at $\theta=90^{\circ}$ we obtain the well established results of references \cite{ketterle1999,ketterle2003}. Fan patterns are observed in the long-pulse regime and X-shape patterns are observed in the short-pulse regime. In superradiant scattering from an elongated BEC, emission of scattered light occurs mainly along the long axis because the gains are maximum along this direction, leading to two end-fire optical modes. Because of this superradiant Rayleigh scattering, the atoms of the BEC undergo sequential processes of absorption-emission of photons from or in any of the two end-fire modes and incident optical mode, which explains the observed patterns. When $\theta=0^{\circ}$, the similar line-patterns with a clear asymmetry \cite{ljt} are also observed.

\begin{figure}[htbp]
\centering
\includegraphics[width=8.5cm]{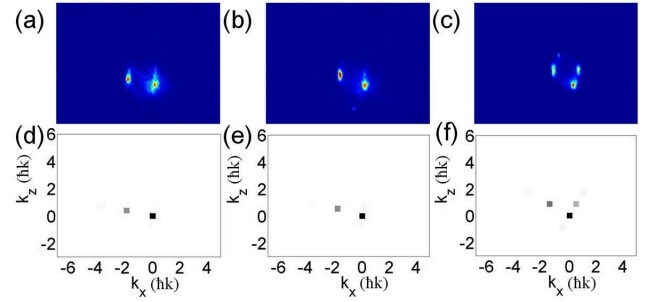}
\caption{(Color online) Superradiant scattering with a long and weak pump beam ($T=200\ \mathrm{\mu s}$, $I=15\ \mathrm{mW/cm^{2}}$) incident at different angles. (a)-(c) Absorption images after 30 ms time of flight. The field of view of each absorption image is 1.9 mm by 1.4 mm. The incident angles are (a) $\theta=24^{\circ}$, (b) $\theta=34^{\circ}$, and (c) $\theta=60^{\circ}$. (d)-(f) Associated calculated atomic side mode distribution patterns with coupling constant $g=1.3\times10^{6}\ \mathrm{s}^{-1}$ and pulse duration $T=200\ \mathrm{\mu s}$.}
\end{figure}

In order to study the asymmetry of superradiant scattering at other angles, the BEC is illuminated with a pump beam incident at a variable angle $\theta$ between $90^{\circ}$ and $0^{\circ}$ in both the long-pulse and the short-pulse regimes. As we will see, superradiant light scattering is still predominant along the long axis of the cigar-shaped BEC at any incident angle. As in previous experiments, atomic side modes can then be conveniently denoted in momentum space. Each side mode is labeled with a pair of integers $(n,m)$ standing for the acquired momentum $\mathbf{q}_{n,m}=\hbar(n\vec{k}_{l}+m\vec{k})$ after the scattering process where $n$ denotes the incident beam direction and $m$ is the long axis of the condensate, $\vec{k}_{l}$ and $\vec{k}$ denoting the wave vectors of the incident light and end-fire modes respectively [cf. Fig. 1(b)]. The momentum of the side mode then takes the form
\begin{equation}
       \mathbf{q}_{n,m}=\hbar[nk_{l}\sin\theta\hat{\mathbf{e}}_x+(nk_{l}\cos\theta+mk)\hat{\mathbf{e}}_z].
\end{equation}

First the BEC is illuminated by a weak pump beam in the long-pulse regime, typically with duration $T=200\ \mathrm{\mu s}$ and intensity $I=15\ \mathrm{mW/cm^{2}}$. At $\theta=24^{\circ}$, only the $(1,1)$ mode is observed, see Fig. 2(a). When $\theta$ is changed to $34^{\circ}$, the $(1,1)$ mode is observed but the $(1,-1)$ mode is still unobservable [cf. Fig. 2(b)]. If $\theta$ is set to $60^{\circ}$, both the $(1,1)$ mode and the $(1,-1)$ mode are populated, see Fig. 2(c). From those time of flight images, we measure the momentum of each mode. For $\theta=24^{\circ}$, $\theta=34^{\circ}$, and $\theta=60^{\circ}$, the moduli of momenta are $1.85\hbar k$, $1.81\hbar k$, and $1.6\hbar k$, respectively. And the recoiling atoms propagate at angles of $12.2^{\circ}$, $18.5^{\circ}$, and $33^{\circ}$ with respect to the $\mathbf{\hat{z}}$ axis, respectively. For the $(1,-1)$ side mode at $\theta=60^{\circ}$, the modulus of momentum is $0.97\hbar k$ and the recoiling atoms propagate at an angle of $109^{\circ}$ with respect to the $\mathbf{\hat{z}}$ axis. The measured momenta are in good quantitative agreement with Eq. (1). It may be experimental evidence that photons are predominantly emitted along the long axis of a cigar-shaped BEC at any incident angle. Comparing these absorption images, we note that, at a low laser intensity, the $(1,-1)$ becomes populated only at a large angle and the $(1,1)$ mode looks a little bit stronger at a small angle $\theta$. It is important to notice that superradiance only occurs when the gain of the side mode is above a certain threshold as discussed in \cite{ketterle1999}. Because superradiance into side mode $(1,1)$ dominates, it can be assumed only the gain for this mode is larger than the threshold for superradiance at a small angle $\theta$ and relatively low intensities.

\begin{figure}[htbp]
\centering
\includegraphics[width=8.5cm]{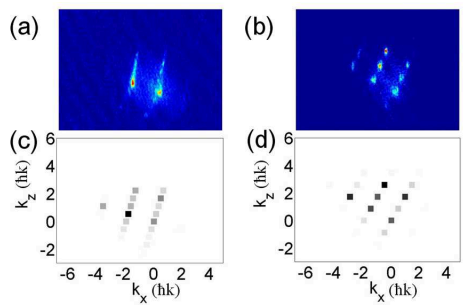}
\caption{(Color online) Superradiant scattering with a long and high-power pump beam ($T=200\ \mathrm{\mu s}$, $I=40\ \mathrm{mW/cm^{2}}$) incident at different angles. (a) and (b) Absorption images after 30 ms time of flight. The field of view of each absorption image is 1.9 mm by 1.4 mm. The incident angles are (a) $\theta=34^{\circ}$ and (b) $\theta=60^{\circ}$. (c) and (d) Associated calculated atomic side mode distribution patterns with coupling constant $g=2.3\times10^{6}\ \mathrm{s}^{-1}$ and pulse duration $T=200\ \mathrm{\mu s}$.}
\end{figure}

Then the BEC is illuminated by a high-power pump beam in the long-pulse regime, typically duration $T=200\ \mathrm{\mu s}$ and intensity $I=40\ \mathrm{mW/cm^{2}}$. At $\theta=34^{\circ}$, many high-order side modes are observed and they are very close to each other [see Fig. 3(a)]. If the angle $\theta$ is changed to $60^{\circ}$, many high-order side modes are also observed but are well separated [cf. Fig. 3(b)]. Previously, we found side mode $(1,1)$ was the first to appear at low pumping intensity. When comparing the two absorption images of Fig. 3 at higher pump intensities, we find that, at $\theta=34^{\circ}$, the superradiant gain for side mode $(2,2)$ is unexpectedly smaller than for other higher order modes with $m<2$ that are seen on the right and top of modes $(0,0)$ and $(1,1)$. This can be explained by the fact the gain of those side modes are above a certain threshold for sequential superradiance, which, as we will explain it further, is due to a rapid but asymmetrical build-up of the end-fire modes.

Up to now, we have only studied the long pulse regime. For exploring the short pulse regime, the BEC is illuminated by a short and strong pump pulse, with $T=30\ \mathrm{\mu s}$, $I=130\ \mathrm{mW/cm^{2}}$ typically. In this regime, both forward and backward recoiling patterns are observed, as can be seen in Fig. 4(a), (b) and (c). At $\theta=24^{\circ}$ and $\theta=34^{\circ}$, the right part of the forward pattern and the left part of the backward pattern [modes $(1,-1)$ and $(-1,1)$ respectively] are unobservable. At $\theta=60^{\circ}$, the left part of the forward pattern looks stronger than the right part. The side mode $(2,0)$ does not become populated due to the missing spatial overlap between atomic modes and relevant light fields, as discussed in \cite{Zobay}.

\begin{figure}[htbp]
\centering
\includegraphics[width=8.5cm]{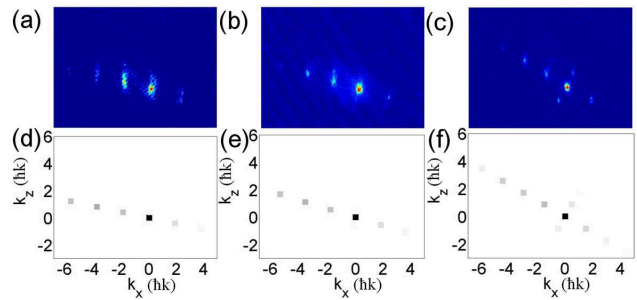}
\caption{(Color online) Superradiant scattering with a short and strong pump beam ($T=30\ \mathrm{\mu s}$, $I=130\ \mathrm{mW/cm^{2}}$) incident at different angles. (a)-(c) Absorption images after 30 ms time of flight. The field of view of each absorption image is 1.9 mm by 1.4 mm. The incident angles are (a) $\theta=24^{\circ}$, (b) $\theta=34^{\circ}$, and (c) $\theta=60^{\circ}$. (d)-(f) Associated calculated atomic side mode distribution patterns with coupling constant $g=4.6\times10^{6}\ \mathrm{s}^{-1}$ and pulse duration $T=30\ \mathrm{\mu s}$.}
\end{figure}

\section{THEORETICAL ANALYSIS AND NUMERICAL SIMULATION}
In the present work, all the absorption images clearly tell us that the recoiling patterns are strongly dependent on the incident angle $\theta$. In order to understand the observed phenomenon, we theoretically model the experiment. An elongated BEC is considered with length $L$ oriented along the $\mathbf{\hat{z}}$ axis. A linearly polarized stimulating laser pulse propagates at a variable angle $\theta$ with respect to the $\mathbf{\hat{z}}$ axis, $\mathbf{E}_l=[\mathcal{E}_le^{-i(\omega_lt-k_lx\sin\theta-k_lz\cos\theta)}\hat{\mathbf{e}}_y+c.c.]/2$. The end-fire modes are $\mathbf{E}_\pm=\mathcal{E}_\pm e^{-i(\omega_\pm t\mp kz})\hat{\mathbf{e}}_y+c.c.$. Taking into account the angular term effect, the Maxwell-Schr\"{o}dinger equations \cite{Zobay} which describe the coupled dynamics of the condensate wave function $\mathbf{\Psi}(\mathbf{r},t)$ and the total electric field $\mathbf{E}(\mathbf{r},t)$ read
\begin{equation}
       \mathrm{i}\hbar\frac{\partial}{\partial t}\mathbf{\Psi} =
       -\frac{\hbar^2}{2M}\nabla^2\mathbf{\Psi}+\frac{(\mathbf{d}\cdot\mathbf{E}^{(-)})(\mathbf{d}\cdot\mathbf{E}^{(+)})}{\hbar\delta}\mathbf{\Psi},
\end{equation}
\begin{equation}
       \frac{\partial^2\mathbf{E}^{(\pm)}}{\partial t^2}=c^2\nabla^2\mathbf{E}^{(\pm)}-\frac{1}{\varepsilon_0}\frac{\partial^2\mathbf{P^{(\pm)}}}{\partial t^2},
\end{equation}
where $\mathbf{d}$ is the atomic dipole moment, $\delta$ is the detuning of the pump laser frequency from resonance. The polarization is given by $\mathbf{P^{(+)}}(\mathbf{r},t)=-\mathbf{d}|\mathbf{\Psi}(\mathbf{r},t)|^2\mathbf{d}\cdot\mathbf{E}^{(+)}/\hbar\delta$, with $\mathbf{P}^{(-)}=\mathbf{P}^{(+)*}$. The energy of atoms in side mode $(n,m)$ is $\hbar\omega_{n,m}=\hbar^{2}(n^{2}k^{2}_{l}+m^{2}k^{2}+2nmk_{l}k\cos\theta)/2M$, where $M$ is the atomic mass. Given that $\omega_{n,m}\ll\omega_{l}$ and because of energy conservation during Rayleigh scattering, we approximately have $k_{l}\approx k$. Thus the side mode frequency is approximately given by $\omega_{n,m}\approx(n^{2}+m^{2}+2nm\cos\theta)\omega_{r}$, where $\omega_{r}=\hbar k^{2}_{l}/2M$ is the recoil frequency. We assume the fields can be decomposed as
\begin{equation}
       \mathbf{\Psi}(\mathbf{r},t) = \sum_{n,m}\frac{\mathbf{\Psi}_{n,m}(z,t)}{\sqrt{A}}e^{-i[\omega_{n,m}t-nk_{l}x\sin\theta-(nk_{l}\cos\theta+mk)z]},
\end{equation}
\begin{eqnarray}
\mathbf{E}^{(+)} = && \frac{\mathcal{E}_l}{2}e^{-i(\omega_lt-k_lx\sin\theta-k_lz\cos\theta)}\hat{\mathbf{e}}_y+\mathcal{E}_{+}e^{-i(\omega_{+}t-kz)}\hat{\mathbf{e}}_y\nonumber\\
&&+\mathcal{E}_{-}e^{-i(\omega_{-}t+kz)}\hat{\mathbf{e}}_y,
\end{eqnarray}
$\mathbf{E}^{(-)}=\mathbf{E}^{(+)*}$, where $A$ is the cross-sectional area of the BEC perpendicular to the direction of the light emission. We solve the Maxwell-Schr\"{o}dinger equations under the slowly-varying-envelope approximation (SVEA) and disregard photon exchange between end-fire modes. Introducing the dimensionless time $\tau=2\omega_{r}t$ and length $\xi=k_{l}z$, we get the following dynamical equations
\begin{eqnarray}
	i\frac{\partial\mathbf{\Psi}_{nm}}{\partial \tau}= && -\frac{1}{2}\frac{\partial^2\mathbf{\Psi}_{nm}}{\partial \xi^2}-i(m+n\cos\theta)\frac{\partial\mathbf{\Psi}_{nm}}{\partial \xi}\nonumber\\
&&+\kappa[e^{*}_{+}(\xi,\tau)\mathbf{\Psi}_{n-1,m+1}(\xi,\tau)e^{i(n-m-2)(1-\cos\theta)\tau}\nonumber\\
&&+e^{*}_{-}(\xi,\tau)\mathbf{\Psi}_{n-1,m-1}(\xi,\tau)e^{i(n+m-2)(1+\cos\theta)\tau}\nonumber\\
&&+e_{+}(\xi,\tau)\mathbf{\Psi}_{n+1,m-1}(\xi,\tau)e^{-i(n-m)(1-\cos\theta)\tau}\nonumber\\
&&+e_{-}(\xi,\tau)\mathbf{\Psi}_{n+1,m+1}(\xi,\tau)e^{-i(n+m)(1+\cos\theta)\tau}]
\end{eqnarray}
with
$$e_{\pm}=\frac{1}{\mathcal{E}_{\pm}}\sqrt{\frac{\hbar\omega k_{l}}{2\varepsilon_{0}A}},$$
$$g=\frac{|\mathbf{d}|^{2}\mathcal{E}_{l}}{2\hbar^{2}\delta}\sqrt{\frac{\hbar\omega_{l}}{2\varepsilon_{0}AL}}=\frac{3\pi c^{2}}{2\omega^{2} \delta}\sqrt{\frac{cI}{AL\hbar\omega_{l}}},$$
and $\kappa=g\sqrt{k_{l}L}/2\omega_{r}$. Neglecting retardation effects, the envelope functions $e_{\pm}$ of the end-fire modes are given by
\begin{eqnarray}
e_{+}(\xi,\tau)= && - i\frac{\kappa}{\chi}\int^{\xi}_{-\infty}d\xi^{'}\sum_{n,m}e^{i(n- m)(1- \cos\theta)\tau}\nonumber\\
&&\times\psi_{nm}(\xi^{'},\tau)\mathbf{\psi}^{*}_{n+1,m-1}(\xi^{'},\tau),
\end{eqnarray}
\begin{eqnarray}
e_{-}(\xi,\tau)= && - i\frac{\kappa}{\chi}\int^{\infty}_{\xi}d\xi^{'}\sum_{n,m}e^{i(n+ m)(1+ \cos\theta)\tau}\nonumber\\
&&\times\psi_{nm}(\xi^{'},\tau)\mathbf{\psi}^{*}_{n+1,m+1}(\xi^{'},\tau),
\end{eqnarray}
where $\chi=ck_{l}/2\omega_{r}$. Using the coupling constant $g$ calculated with corresponding experimental parameters, the calculated atomic side mode distribution patterns with the same initial seed equal to one atom, are shown in the lower parts of Figs. 2, 3, and 4. As can be seen, they are in quantitative agreement with the experimental results. It is important to notice that Eqs. (6), (7) and (8) naturally imply the angular term $\theta$ strongly influences the cooperative scattering of light and atoms in the BEC. Distributions of the envelope functions $e_{+}$ and $e_{-}$ of the end-fire modes as a function of the incident angle $\theta$ and $z$ coordinate are simulated at given parameters in long-pulse regime, see Fig. 5(a). The $|e_{-}|$ modulus seems to keep the same level when changing the incident angle $\theta$. The $|e_{+}|$ modulus is at the same level with $|e_{-}|$ when the angle $\theta$ is very close to $90^{\circ}$. When decreasing the angle $\theta$, $|e_{+}|$ becomes stronger, but localized on the one-end side of the BEC. If decreasing the angle $\theta$ close to $0^{\circ}$, $|e_{+}|$ drops rapidly to very small values. The angular dependence of the side modes populations is also simulated, see Fig. 5(b). These numerical simulations fit the main features of the experimental results.

\begin{figure}[htbp]
\centering
\includegraphics[width=8.5cm]{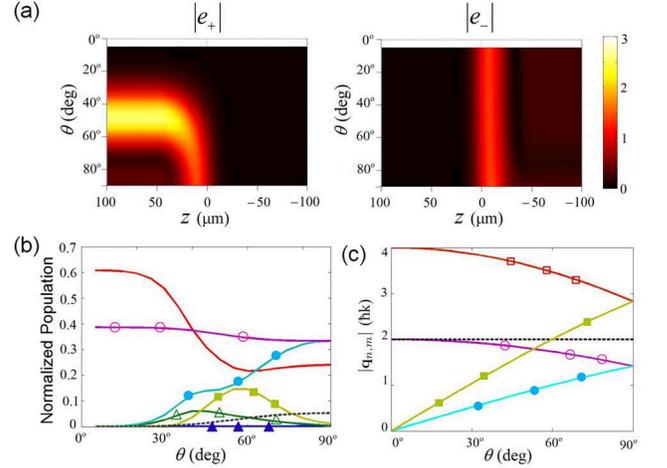}
\caption{(Color online) Theoretical calculations for the angular spectroscopy of superradiant scattering from a BEC. (a) Distributions of the envelope functions $e_{+}$ and $e_{-}$ of the end-fire modes as a function of the incident angle $\theta$ and $z$ coordinate. (b) Angular dependence of the side mode population versus $\theta$. $(0,0)$(solid line), $(1,1)$(open circles), $(1,-1)$(filled circles), $(-1,1)$(open triangles), $(-1,-1)$(filled triangles), $(2,-2)$(filled squares), $(2,0)$(dashed line). These figures are given for a $g=1.5\times10^{6}\ \mathrm{s}^{-1}$ coupling constant and pulse duration $T=200\ \mathrm{\mu s}$. (c) Angular variation of the side mode momenta norms $|\mathbf{q}_{n,m}|$. $(1,-1)$(filled circles), $(1,1)$(open circles), $(2,-2)$(filled squares), $(2,2)$(open squares), $(2,0)$(dashed line).}
\end{figure}

Though exact understanding of the evolution of the fields requires to look at their spatial distributions and time evolutions quite in detail, some reasonable explanations can be proposed. Because of the relatively poor overlap of the $|e_{+}|$ end-fire mode with the BEC for $\theta \neq 90^{\circ}$, only a weak superradiance effect is observed at low intensities. Especially, at small angles $\theta$, the right part of the recoiling pattern is unobservable and the pattern looks like a line-pattern in the experiment. Increasing the angle $\theta$, sequential scattering can be very strong, but populates mainly modes with $m<2$ on the right and top of modes $(0,0)$ and $(1,1)$. When the angle $\theta$ is close to $90^{\circ}$, a nearly symmetric pattern is observable and the pattern is similar to the fan-like or X-shape pattern.
To understand this transition, one has to look at the time evolution of the spatial distributions of the different modes. Indeed, the initial build-up of side mode $(1,1)$ tends to deplete the BEC close to its center, where end-fire mode $|e_{-}|$ is the highest. Because of this localization of the end-fire mode $|e_{-}|$, the gain for further growth of the $(1,1)$ order is limited. On the contrary, the end-fire mode $|e_{+}|$ seems to grow very rapidly on the other side of the condensate, especially for $\theta$ in between $\sim 30^\circ$ and $\sim 70^\circ$, which leads to a concomitant growth of the side modes on the top right of the BEC.

From the previous theoretical study, we have obtained a good understanding of superradiant scattering as a function of the incident angle. It is now also interesting to notice the angular dependence of $|\mathbf{q}_{n,m}|$ in Eq. (1). In the case of $|\mathbf{q}_{n,\pm n}|$, at $\theta=0^{\circ}$, $|\mathbf{q}_{n,n}|=2n\hbar k_{l}$ and $|\mathbf{q}_{n,-n}|=0\hbar k_{l}$. Increasing the angle $\theta$, $|\mathbf{q}_{n,n}|$ decreases and $|\mathbf{q}_{n,-n}|$ increases. Finally they reach the same value $\sqrt{2}n\hbar k_{l}$ at $\theta=90^{\circ}$. In the case of $|\mathbf{q}_{2n,0}|$, the modulus of $|\mathbf{q}_{2n,0}|$ is $2nk_{l}$ and it is independent of the angle $\theta$. For the sake of illustration, the angular dependence of $|\mathbf{q}_{n,m}|$ is plotted in Fig. 5(c). It is clearly shown that the momenta can continuously cover a large range when changing the angle $\theta$. Extrapolating our observations, angular superradiance may be a useful technique for obtaining very broad optional momenta.

\section{CONCLUSION}
In summary, we have presented an investigation of the angular spectroscopy of superradiance from a BEC. By exploring the incident angle dependence of superradiant Rayleigh scattering, we have shown the evolution of the recoil pattern between symmetric superradiance and asymmetric superradiance depends much on the experimental conditions. We have experimentally demonstrated this transition occurs at different angles depending on the pumping intensity for example. Specific recoiling atomic patterns have been observed in this study for the first time because the spatial distributions of side modes and end-fire modes are strongly dependent on the incident angle. Taking into account the angle of incidence, calculated atomic recoiling patterns and numerical simulations of end-fire modes are consistent with experimental results. We think this work will be important in the quest for acquiring a full-knowledge of superradiant scattering processes in Bose-Einstein condensates.

\begin{acknowledgments}
This work received support from the National Fundamental Research Program of China under Grant No. 2011CB921501, the National Natural Science Foundation of China under Grant No. 61027016, No. 61078026, No. 10874008 and No. 10934010.
\end{acknowledgments}

\end{document}